\documentclass{mn2e}
\usepackage{psfig}
\def    \kms    {\rm km\ s^{-1}}

\def    \modot  {\rm M_{\odot}}

\def\pasp{{Publ. Astron. Soc. Pac.}}

\def\apj{{Astrophys. J.}}

\def\mnras{{Mon. Not. R. Astron. Soc.}}
\def\apjs{{Astrophys. J., Suppl. Ser.}}
\def\aj{{Astron. J.}}

\def\aaps{{Astron. Astrophys. Suppl. Ser.}}

\begin{document}

\title[Nearest O3 Star]   
{The Nearest Star of spectral type O3: A Component of the Multiple System 
HD 150136}

\author[V.S. Niemela \& R.C. Gamen]{ V.S. Niemela $^1$
\thanks {Member of Carrera del Investigador, CIC--BA, Argentina}, 
and R.C. Gamen $^2$ 
\thanks {Visiting Astronomer, CASLEO, San Juan, Argentina }
\\                                                
$^1$ Facultad de Cs. Astron\'omicas y Geof\'{\i}sicas,
Universidad Nacional de La Plata, Paseo del Bosque s/n, 1900 La Plata, Argentina;\\ virpi@fcaglp.unlp.edu.ar \\ 
$^2$ Departamento de F\'\i sica, Universidad de La Serena, Benavente 980,
La Serena, Chile;\\ rgamen@dfuls.cl\\ 
}                    
\maketitle
 
\begin {abstract}
From radial velocities determined in high signal-to-noise digital 
spectra, we report the discovery that the brightest component of
the binary system HD 150136 is of spectral type O3. We also
present the first double-lined orbital solution for this binary.
Our radial velocities confirm the previously published spectroscopic orbital 
period of 2.6 days.  He\,{\sc ii} absorptions appear double at quadratures,
but single lines of N\,{\sc v} and N\,{\sc iv} visible in our spectra define 
a radial velocity orbit of higher semi-amplitude for the primary component
than do the  He\,{\sc ii} lines. 
From our orbital analysis, we obtain minimum masses for the binary components
of 27 and 18$\modot$.
The neutral He absorptions apparently do not follow the orbital motion of 
any of the binary components, thus they most probably arise in a third star 
in the system. 

\end{abstract}
 
   
\section { Introduction}
HD~150136 ($\alpha$(2000)=16h41m24s;  $\delta$(2000)=-48$\degr$45'47'';  V=5.5), 
together with its 
close companion HD 150135, are the brightest
stars of the open cluster NGC~6193 in the nucleus of the  Ara OB1 association.
Their spectra have been classified by Walborn (1972) as O5\,III:n(f) 
and O6.5V((f)), respectively. HD 150136 appears as a double lined binary in the
compilation of binary stars of spectral type O (Garmany et al. 1980), 
with a period of 2.7 days and mass ratio of 1.8, but the orbit 
has not been published yet.  Arnal et al. (1988) in their 
search for short period binaries in the open cluster NGC 6193, published 
a single-lined orbit with a period of 2.6 days.

HD~150136 is also the brightest star of a visual multiple system with
6 components within a radius of about 20 arcseconds. The closest component
to  HD~150136 is a 9th magnitude star (component B of the system) at 
1.6 arcseconds (cf. Mason et al. 1998). HD~150135 (V=6.7, component C) is
the second brightest star in this group at a separation of 9.6 arcseconds from
HD~150136.

In this paper, we report the first double-lined orbital solution for 
HD~150136 based on medium resolution digital high S/N spectral images. 
Furthermore, our spectra and radial velocities suggest that the primary 
star has a very early spectral type O3. At the distance of AraOB1, 
about 1300pc (cf. Herbst \& Havlen 1977), this
star then becomes the nearest star of this early spectral type.

\section {Observations}
We have obtained 19 digital spectral images of HD 150136 with the REOSC
Cassegrain spectrograph attached to the 2.1m reflector at the Complejo
Astronomico El Leoncito (CASLEO\footnote{ CASLEO is operated under agreement 
between CONICET, SECYT, and the National Universities of La Plata, 
C\'ordoba and San Juan, Argentina.}) in San Juan, Argentina. The spectra were
observed in May and June 2004. A TEK 1024 $\times$ 1024 pixels CCD was used as
detector. The reciprocal dispersion of the spectra is 1.8\AA\ $/$ pixel. 
Exposure times were between 1.5 and 5 minutes, resulting in spectra of 
signal-to-noise ratio between 100 and 150. The wavelength range covered in our
spectral images is from  3800 to 5500\AA\ . Comparison lamp spectra 
of CuAr
were observed immediately after or before at the same telescope position as 
the stellar images. Flat-field and bias frames were also observed each night.
Onedimensional spectra were extracted from the twodimensional spectral images
using IRAF routines.

\section {Results and their discussion}

\subsection{The spectrum of HD 150136}
Along with the characteristics of the previously determined spectral type
of O5\,III:n(f), i.e. strong absorption lines of H and  He\,{\sc ii}, fainter 
absorptions
of He\,{\sc i}, and N\,{\sc iii} emission at  4634-40\AA\, our spectra 
of HD~150136 
show also absorption lines of N\,{\sc v}  4603-19 \AA\ and N\,{\sc iv}
 5203~\AA\, as well as the emission line of N\,{\sc iv}  4058~\AA\ .  
These lines are the distinctive ones observed in 
earlier O2-3 type spectra, (cf. Walborn et al. 2002; Gamen \& Niemela 2002).
One of our spectra of HD~150136 is illustrated in figure 1.
The simultaneous presence of He\,{\sc i} and N\,{\sc v} absorptions in 
the spectrum
clearly pointed to a composite spectrum and to the possibility that the 
brighter binary component
is of earlier spectral type than O5. We therefore decided to study the
radial velocities of the lines in the spectrum of HD~150136.

\begin{figure*}
\vspace {15cm}
\includegraphics{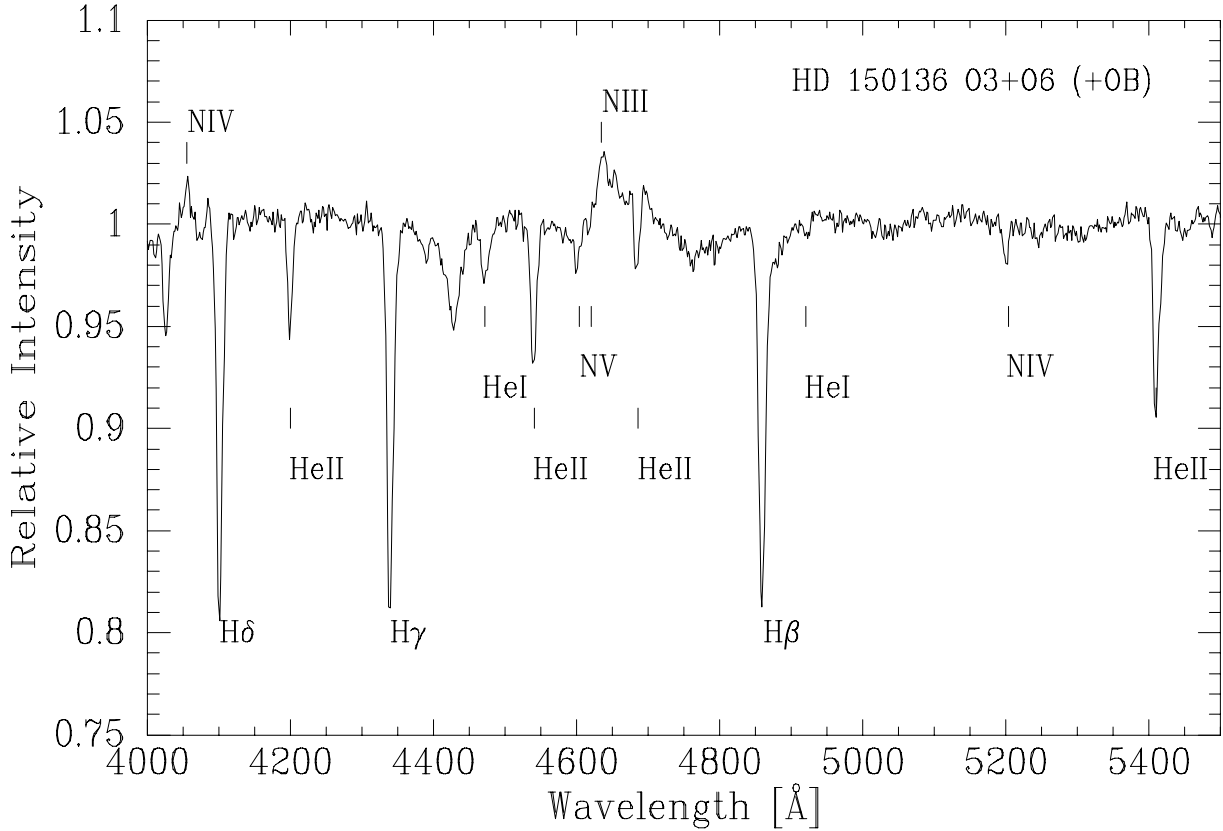}
\caption {Continuum rectified spectrum of HD~150136 obtained at CASLEO in 2004,
May. Absorption lines identified below the continuum are 
H$\delta$\,$\lambda$4102,
H$\gamma$\,$\lambda$4340, H$\beta$\,$\lambda$4862, He\,{\sc ii} 
\,$\lambda\lambda$ 4200, 4541, 4686, 5411, 
He\,{\sc i}\,$\lambda\lambda$4471, 4922,  
N\,{\sc v}\,$\lambda\lambda$4604, 4620, and N\,{\sc iv}\,$\lambda$5403. The
emission lines identified above the continuum are N\,{\sc iv}\,$\lambda$4058
and N\,{\sc iii}\,$\lambda\lambda$4634--4640--4642.}
\label{fig01}
\end{figure*}

\begin{table*}
\caption{ Journal of spectral observations of HD~150136}
\begin{center}
\begin{tabular}{ccccc}
\hline
HJD & Phase & \multicolumn{3}{c}{Heliocentric Radial Velocity ($\kms$)}\\
\cline{3-5}
   & & Primary (N {\sc v, iv}) & Primary (He {\sc ii})&  Secondary\\
\noalign{\smallskip}\hline\noalign{\smallskip}
    3146.856 &     0.87    & 152        &            71 & 242     \\
    3150.849 &     0.37    &-211        &            -158 &236    \\
    3151.769 &     0.71    &216         &            101 &-345     \\
    3153.809 &     0.48    &-65         &            -64 &        \\
    3154.769 &     0.84    &156         &            106 &-281    \\
    3155.761 &     0.21    &-235        &            -186 &316    \\
    3156.707 &     0.57    &19          &            -17 & \\
&     &   &  \\
    3169.640 &     0.43    &-133        &            -109 &    \\
    3169.804 &     0.49    &-57         &             -62 & \\
    3170.501 &     0.75    &212         &            101 &-358 \\
    3170.505 &     0.75    &205         &            114 &-345 \\
    3170.548 &     0.77    &181         &            120 &-353  \\
 3171.510 &     0.13    &-165        &             -82 &    \\
    3171.614 &     0.17    &-176        &            -156 &232  \\
    3171.705 &     0.20    &-214        &            -151 &246 \\
    3171.850 &     0.26    &-237        &            -161 &324   \\
    3171.873 &     0.27    &-215        &            -144 &304   \\
    3172.666 &     0.56    & 37         &             5   &   \\
    3172.824 &     0.62    & 143        &            34 &  \\
\noalign{\smallskip}\hline\noalign{\smallskip}

\multicolumn{5}{l}{HJD $=$ Heliocentric Julian 
Date\,$-\,2\,450\,000$\,d}\\
\multicolumn{5}{l}{Phases were calculated according to T$_0$ = 
2\,453\,171.2+2.662\,E}\\

\end{tabular}
\end{center}
\label{tab01}
\end{table*}

\subsection{The radial velocity orbit}
Radial velocities were determined fitting gaussian profiles to the spectral
lines within the IRAF routine SPLOT.  He\,{\sc ii} absorptions, and to a lesser
extent also those of Hydrogen Balmer lines, have the appearance of double
lines of unequal components in several of our spectra. We did not use the
H absorption lines in our radial velocity study. Radial velocities
for He\,{\sc ii} lines were determined using the deblending routine of SPLOT. 
The absorption lines of N\,{\sc v} and N\,{\sc iv}, as well as the N\,{\sc iv}
emission, appear single in all of our spectra, and we 
used the radial velocities of these lines to determine an approximate orbital
period. The journal of our radial velocity observations is presented in Table~1,
and spectra corresponding to approximately opposite orbital phases ($\phi$=0.21
and $\phi$=0.71) are depicted
in Figure~2, where the lines of both components can be appreciated.
 
We introduced the values of N\,{\sc v} and N\,{\sc iv} radial velocities in the 
period search routine of Lafler \& Kinmann (1965). As can be seen from the 
values listed in Table~1, the radial velocities from our spectra
show large variations from one night to another, but smaller variations 
between observations obtained during the same night, thus indicating a binary 
period of a few days. We therefore searched for periods between 1 and 5 days. 
The best period we found was 2.65 days, almost identical to the values found 
previously,
namely 2.7 days and 2.6 days (cf. Garmany et al. 1980, Arnal et al. 1988). 

The value of the period we found was then introduced as an initial value to the 
routine for defining the orbital elements of the binary. To this end we 
used an improved version\footnote {available upon request from 
$ftp://lilen.fcaglp.unlp.edu.ar/pub/fede/gbart-0.1-41.tar.gz$} 
of the program originally published by Bertiau \& Grobben (1969).  
The radial velocity orbit of the primary star 
was derived using the mean of the radial velocities of N\,{\sc v} 
 4603-19\AA\ and N\,{\sc iv}  5203 \AA\ absorptions and 
the N\,{\sc iv} 4058\AA\ emission line, 
since these lines likely originate only in the brighter component. The orbit 
of the secondary component was determined from the radial velocities of the 
 He\,{\sc ii} absorptions observed when the components appeared most separated.
To a first approximation, the radial velocity orbit appears to be circular,
as we obtained an orbital eccentricity of 0.03 $\pm$ 0.02. Therefore we have
fitted circular orbits to our radial velocity observations.
The orbital elements are listed in Table~2. These orbital elements should be
considered as preliminary values, because due to the rather limited 
wavelength resolution of our spectra, the separation of the spectral lines 
of the secondary component may not be adequate. 
The upper panel in Figure 2 illustrates the double-lined radial velocity
orbit of HD~150136.

The minimum masses that we find for the binary components are moderately high
(cf. Table~2).
However, HD~150136 is not known as an eclipsing binary, and from Hipparcos
photometry Marchenko et al. (1998) find a dispersion of 0.03 mag, with no
trace of periodicity. Thus we may expect the orbital inclination to be 
rather low. An inclination of 50 $\degr$ would result in masses of about
60$\modot$ and 40$\modot$ for the O3 and O6 components, respectively.

\begin{figure*}
\vspace {17cm}
\includegraphics{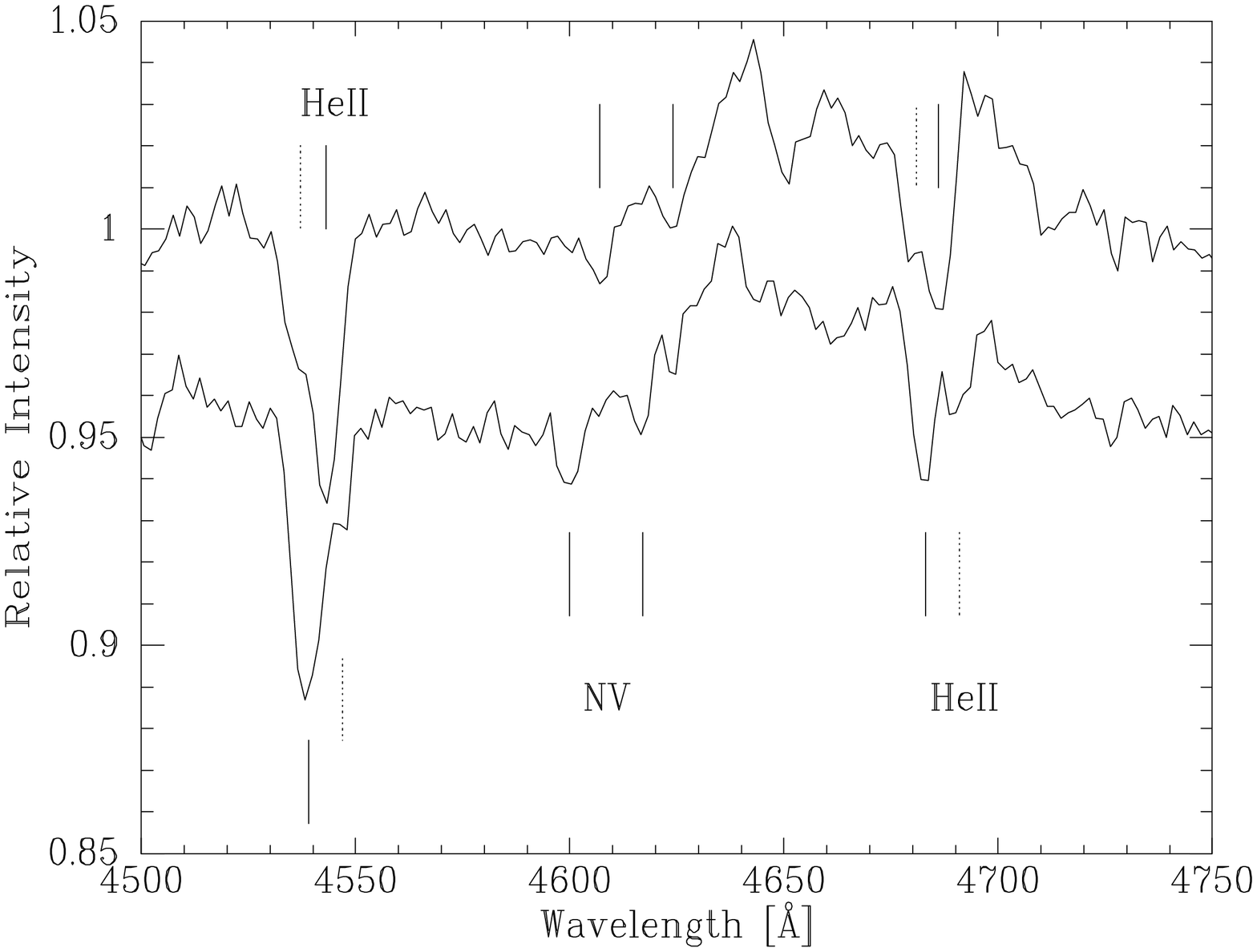}
\caption { Spectra of HD~150136 during opposite orbital phases ($\phi$=0.21
and $\phi$=0.71) are depicted showing the
spectral lines of the two components. Lines of the primary are denoted with
vertical solid lines and those of the secondary component with dotted lines.}
\label{fig02}
\end{figure*}

\begin{figure*}
\vspace {17cm}
\includegraphics{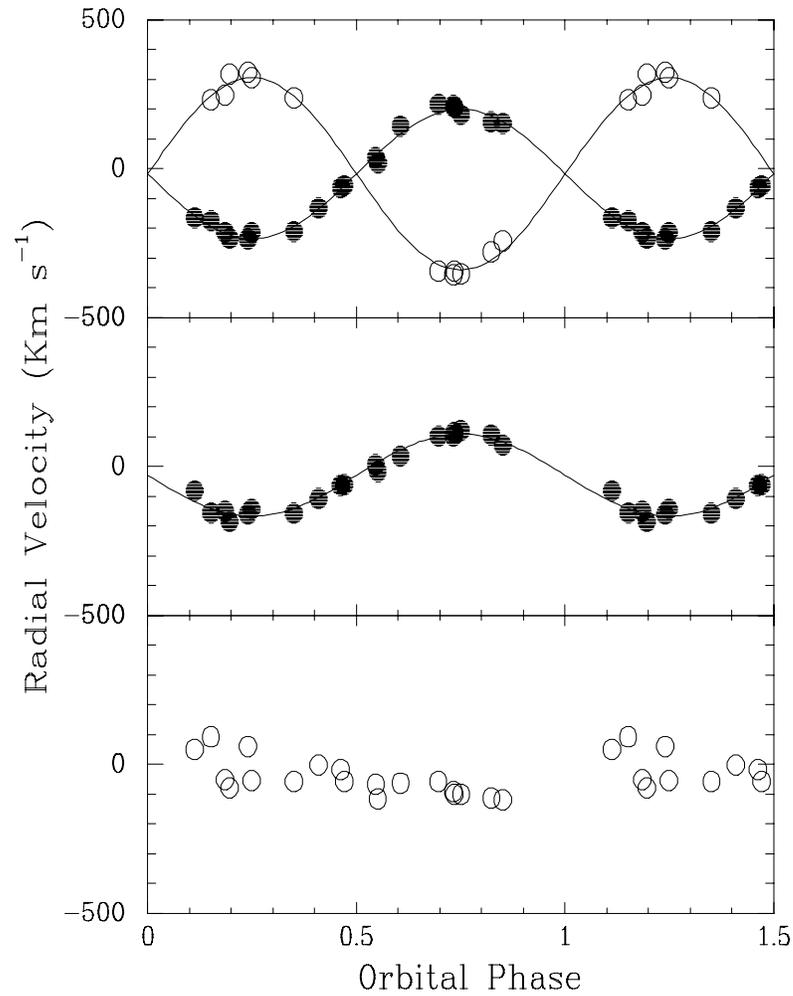}
\caption {{\it Upper panel}: Radial velocity variations of the primary 
component (filled circles) from the N\,{\sc v} and N\,{\sc iv} lines, 
and of the secondary component from He\,{\sc ii} absorptions observed during
maximum separation of the lines (open circles) in the HD~150136 binary system,
phased with the period of 2.662 days. Continuous curves represent the orbital
solutions from Table~2.
{\it Middle panel and lower panel}: Radial velocity variations of the stronger 
components of the He\,{\sc ii} absorptions (filled circles), and 
of the He\,{\sc i} absorptions (open circles) in the 
spectrum of HD~150136,
respectively. All data have been phased with the same ephemeris.}
\label{fig03}
\end{figure*}

\begin{table}
\caption{ Preliminary Circular Orbital Parameters for HD~150136.}
\begin{center}
\begin{tabular}{cccc}
\hline
         & Primary  & Secondary &Primary \\
         & (N\,{\sc v, iv})& & (He\,{\sc ii})\\ 
\hline
&&\\
$a \sin i$ [R$_\odot$]  & 11$\pm$2  & 17$\pm$2 &  \\
$K$ [km s$^{-1}$]       &  217$\pm$4 & 322$\pm$5 & 138$\pm$5 \\
$V_{o}$[km s$^{-1}$]    & -18$\pm$3  & -18$\pm$4 & -30$\pm$5 \\
$M \sin^3i$ [M$_\odot$] & 27$\pm$2   & 18$\pm$2 &  \\
$T_{RVmax}$ [HJD] 2.450.000+&3170.5$\pm$0.2&& \\
$P$ [days] & \multicolumn{2}{c}{2.662$\pm$0.002} &\\
\hline
\end{tabular}
\end{center}
\label{tab02}
\end{table}

The radial velocities of He\,{\sc i} absorption lines apparently do not
follow the orbital motion of any of the binary components. However,
in a few of our spectra, these absorptions show a fainter component to
the red when the secondary of the binary system has its maximum positive
velocity. We therefore assume that the  He\,{\sc i} absorptions mainly originate
in a third star in the system, and thus HD~150136 is a multiple star
similar to Sk-67$\deg$18 in the Large Magellanic Cloud (Niemela, Seggewiss
\& Moffat, 2000). This fact is also supported by the lower amplitude of the
radial velocity variations of the  He\,{\sc ii} absorptions of the primary
component (138$\kms$), as compared with the radial velocity variations 
of N\,{\sc v} and N\,{\sc iv} lines (217$\kms$).

The middle panel and the lower panel in Figure~3 illustrate
the radial velocity variations of the stronger components of He\,{\sc ii}
 and of He\,{\sc i}  absorptions, respectively, phased with the binary period.

A lower amplitude of He\,{\sc ii} absorption radial velocity variations 
probably arises because despite of the possible line blending effects, 
these lines are also blended with those of the third
star in the system. As noted by Andersen (1975) for the diffuse He\,{\sc i}
lines, pair-blending effects are
appreciable on lines which show extensive wings, even if their cores appear
well defined and resolved. This effect probably also affects the He\,{\sc ii}
absorptions. We have tried to correct for pair-blending of He\,{\sc ii}
absorptions in our spectra of HD~150136 using the multiple profile 
deblending routines within IRAF, which fit gaussian, lorentzian, or voigt 
line  profiles with a linear  background. Obviously, to verify the success of 
this procedure in separating the lines of the secondary component, as well 
as to ascertain the contribution of the third star to the stronger components 
of the He\,{\sc ii} absorptions, an improved orbital solution based on high 
resolution observations is needed. However, the high masses of the binary
components for an assumed orbital inclination of 50$\degr$ together with 
the lack of photometric variations (see above), suggest that our preliminary 
values of the semiamplitudes of the radial velocity variations in the 
HD~150136 binary system may not be very far from the true values, unless the 
masses of the binary components are much higher than those determined for 
other stars of similar spectral types (e.g. Massey et al. 2002).

\subsection {The spectral types of the binary components}

As noted in section 3.1, the presence of high ionization N\,{\sc v} and 
N\,{\sc iv}
lines in the spectrum of the primary component points to an early O2-3 spectral
type. The emission and absorption complex of N\,{\sc iii} and He\,{\sc ii} at 
$\sim$ 4634--4690~\AA\ changes its shape with orbital phase. Emission
of He\,{\sc ii} 4686~\AA\ is apparent when the primary is in front of the
system. The spectral classification criteria for O2-4 stars put forward by
Walborn et al. (2002) compare the N\,{\sc iv} emission at 4058~\AA\
with the N\,{\sc iii} emission at  4634--40~\AA\ . In our spectra of
HD~150136 this criterion would indicate a spectral type O3.5. The luminosity
class is difficult to infer, but assuming that the He\,{\sc ii}~4686~\AA\  
PCyg type profile observed when the primary is in front of the system
arises in this component, the primary would appear to be an O3.5~If$^{*}$ type
star.

We have estimated the spectral type of the secondary component in the
HD~150136 binary system from a comparison of the spectral lines of He\,{\sc i}
and He\,{\sc ii}  4471/4541~\AA\ in our spectra when the secondary 
component has its maximum positive radial velocity. This is illustrated 
in Figure~4, where these lines corresponding  to the secondary are denoted 
with dotted vertical lines. In this Figure we note that the 
He\,{\sc ii}  4541\AA\ line corresponding to the secondary is somewhat 
stronger than
He\,{\sc i}  4471\AA\ , indicating a spectral type earlier than O7, where
the relation of these lines becomes unity. Together with the fact that 
He\,{\sc ii}  4686\AA\ line, which is a luminosity indicator in O 
type stars, appears
in absorption (cf. Fig.~2), the spectral type of the secondary component
appears to be O6~V.

\begin{figure*}
\vspace {15cm}
\includegraphics{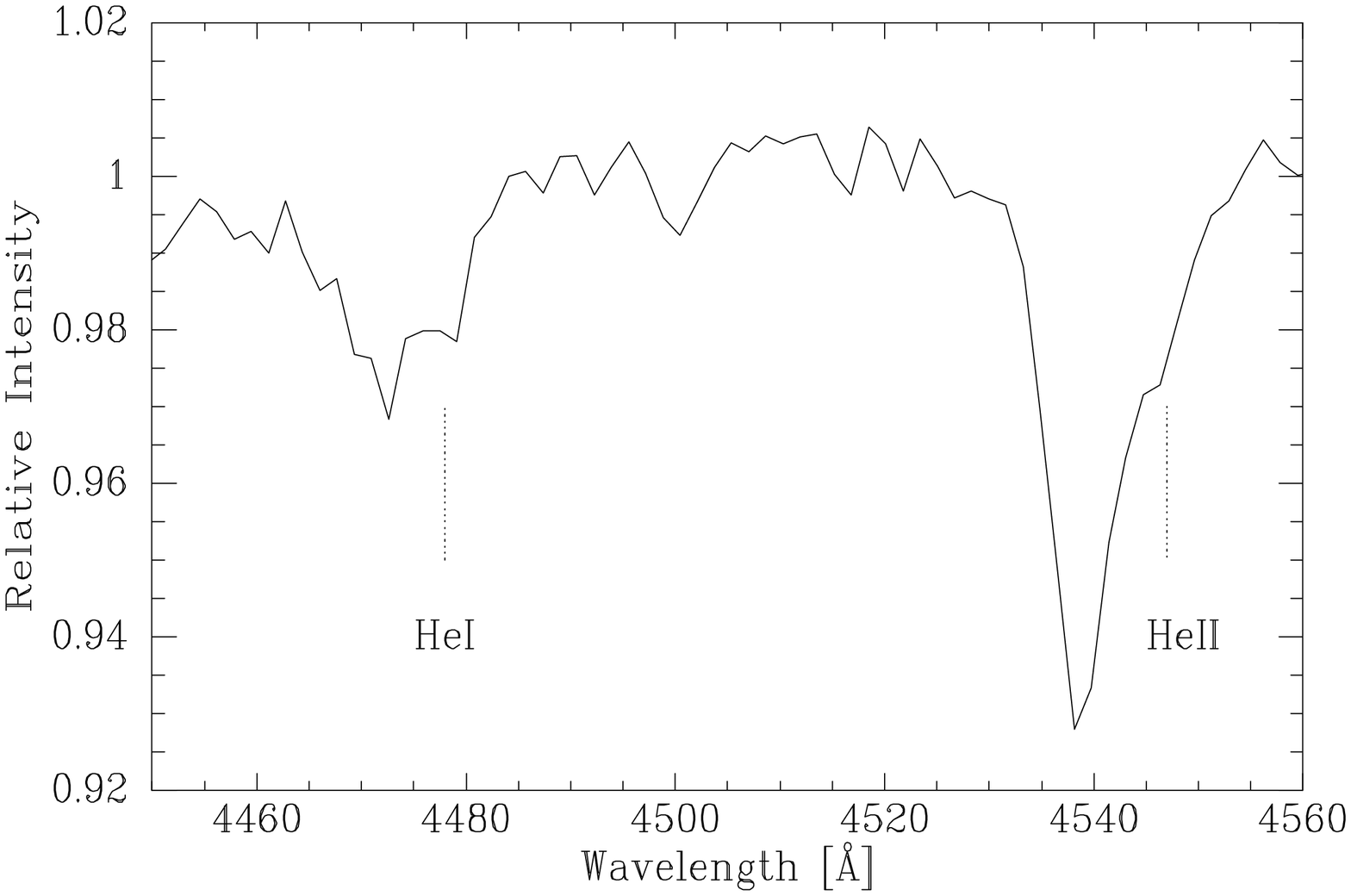}
\caption { Spectrum of HD~150136 during the maximum positive radial velocity
of the secondary component llustrating the relation of absorption lines of 
He\,{\sc i} 4471\AA\ and He\,{\sc ii} 4541\AA\ in the secondary 
component, denoted with vertical dotted lines.} 
\label{fig04}
\end{figure*}

\section {Summary and Conclusions} 
From an analysis of recent high signal-to-noise spectra of HD~150136 we 
find the following:
\begin{enumerate}
\item Radial velocity variations of spectral lines with a period of 2.662 
days are present in our spectra of HD~150136, confirming the previously 
found values of the binary period (Garmany et al. 1980, Arnal et al. 1988).

\item Radial velocity variations of the highest ionization lines, namely
N\,{\sc v} and N\,{\sc iv}, define an orbital motion of rather higher amplitude
for the primary star than the absorptions of He{\sc ii}, 217$\pm$4~$\kms$ vs. 
138$\pm$5~$\kms$, respectively.
\item The radial velocities of the main components of He\,{\sc i} absorptions 
do not follow the
orbital motion of any of the binary components.
Therefore HD~150136 appears to harbour at least another luminous OB type 
component, in addition to the 2.6 day binary system.

\item The primary component of this binary system is of spectral type O3,
noticeably ealier than the spectral classification found in the literature,
namely O5\,III:n(f), which obviously referred to the composite of all 
three spectra present, at a single phase.
 We estimate a spectral type for the secondary component as O6~V.
 \item Previously reported small light variations do not show periodicity, 
thus the orbital inclination is probably low. An inclination of 50 $\degr$ 
would produce masses of 60 and 40 $\modot$ for the O3 and O6 components, 
respectively.
\end{enumerate}
 
HD~150136 is one more example of the increasing number of O stars discovered
to belong to tight
multiple star systems, such as  e.g. CPD -59$\degr$2636 and CPD -59$\degr$ 2603
in the open cluster Trumpler 16 in the Carina Nebula. The former system 
consists of an O7~V+O8~V short period (3.62d) binary  and a fainter O9~V
component (Albacete Colombo et al. 2002); and the latter system similarly of a  
short-period (2.15d) eclipsing O7~V+O9.5~V binary bound to a B0.2~IV star
(Rauw et al. 2001). Another relevant example is HD~167971, a triple system
with a short period (3.32d) eclipsing binary and a more distant  star of
spectral type O8~Ib, which is the most luminous component of this triple system
(Leitherer et al. 1987).

Close triple stars are also found among the Wolf-Rayet stars, thought to be
evolved descendants of O type stars. Notorious examples of these are $\theta$
Mus (=~HD~113904) and HD~5980.  $\theta$~Mus consists of a WC+(O?) short 
period (18d) binary with a wider component which is a more luminous late O 
type supergiant ( cf. Moffat \& Seggewiss 1977). HD~5980  resides in the 
brightest H{\sc ii} region of the Small Magellanic Cloud, and contains a 
Luminous Blue Variable (LBV) which erupted in 1994 ( Barb\'a et al. 1995).
The system consists of two emission line stars 
in an eclipsing binary of short period (19d) and a line of sight O type
companion (cf. Niemela et al. 1999).

Mason et al. (1998) in their astrometric/spectroscopic survey of O stars 
brighter than V$\sim$~8, find a large fraction of primary components in 
close visual systems to be short period spectroscopic binaries. These "hard"
binaries in close triple and multiple systems are clues to our understanding
of massive star formation processes, as well as the formation of runaway stars
by gravitational encounters.

\section{acknowledgements}
We thank the director and staff of CASLEO for the use of their
facilities.
We also acknowledge the use at CASLEO of the CCD and data acquisition
system partly financed by U.S. NSF grant AST-90-15827 to R. M. Rich.
We are grateful to the referee, Dr. Nolan Walborn, for very helpful comments
which improved the presentation of this paper.
We are indebted to Federico Bareilles for the improved version of the
Bertiau \& Grobben program.
This research has received financial support from IALP, CONICET, Argentina.

\end{document}